\documentclass[12]{article}
\usepackage{graphicx}
\usepackage{amssymb}
\begin{document}
\mbox{}\\[1cm]
\begin{center}
{\LARGE \bf Symmetry properties of Penrose type tilings\\[5mm] }
NICOLAE COTFAS\\[3mm]
Faculty of Physics, University of Bucharest, PO Box 76-54, Post Office 76, Bucharest, Romania\\[3mm]
E-mail: ncotfas@yahoo.com  \quad Homepage at \verb#http://fpcm5.fizica.unibuc.ro/~ncotfas# \\[1cm]
\end{center}

\noindent {\small The Penrose tiling is directly related to the atomic structure of certain decagonal quasicrystals and, 
despite its aperiodicity, is highly symmetric.  
It is known that the numbers $1$, $-\tau $, $(-\tau )^2$, $(-\tau )^3$, ... , where $\tau =(1+\sqrt{5})/2$, are scaling factors
of the Penrose tiling. We show that the set of scaling factors is much larger, and for most of them the number of the corresponding 
inflation centers is infinite.}\\[5mm]

\section{Introduction}

The best known and by far most studied aperiodic tiling of plane is the Penrose tiling \cite{Pe,Br}.
This tiling plays an important role in the quasicrystal structure understanding and has some remarkable
mathematical properties. Direct structure observation, using high-resolution transmission electron microscopy, 
shows that the atomic arrangement in the tenfold symmetry plane of certain decagonal quasicrystals can simply 
be interpreted in terms of the Penrose tiling \cite{St,Ab}. Many aspects concerning the mathematical properties have been studied
and are found in the literature; see for example \cite{Ja,Se,Mo}.

In this paper, we present some results concerning the inflation symmetries of the Penrose type tilings. 
Our approach, based on the use of the three projectors corresponding to the usual decomposition of the five-dimensional superspace 
used in the cut and project construction, is an extension of the method used by Katz and Duneau in the description of
self-similarity properties of icosahedral quasicrystals \cite{Ka,Co}. 
There are also two important differences. In the case analysed by Katz and Duneau, the representation
of the icosahedral group in the superspace $\mathbb{R}^6$ is a sum of only two irreducible representations, and the orthogonal 
projection of the lattice $\mathbb{Z}^6$ on the internal subspace is a dense subset. In our case, the representation of the symmetry group $C_5$
in the superspace $\mathbb{R}^5$ is a sum of three real irreducible representations, and the orthogonal projection of the lattice $\mathbb{Z}^5$
on the internal three-dimensional subspace is not a dense subset. 
The vertex set of the Penrose tiling is a multi-component model set \cite{Ba,Mo,Cot}, and we have
adapted the Katz and Duneau method to such a case.

The set $\Lambda $ of inflation factors obtained by using our method is a union of four cut and project sets. 
To our knowledge, the known inflation factors of the Penrose tiling are only those forming the multiplicative semigroup  
$\mathcal{I}=\{ \, (-\tau )^k\ |\ k\!\in \!\mathbb{N}\, \}$.
Since the elements of this semigroup are not uniformly distributed in $\mathbb{R}$ we can not have 
$\Lambda \subset \mathcal{I}$.
We have found new scaling factors and inflation centers, but very probably our list is not exhaustive.
It is known that the inflation factor $\tau $ is directly related to a construction of the Penrose tiling by tile substitutions \cite{Ja,Se}.
We do not know if the new found scaling factors are related to some substitution rules or not.  

The importance of inflation symmetries in quasicrystal physics is comparable with that of translation symmetries
in the description of crystals. A fractal shape of the energy spectrum may be a direct consequence of the self-similarity 
properties of the quasicrystal. A method to study the self-similarity properties of the cut and project sets has been presented 
in \cite{Ma}, but as the authors show at the end of the paper, the method does not apply to the Penrose type tilings.

\section{Penrose type tilings}

In this section we review certain results concerning the Penrose type tilings defined by projection
and introduce certain notations. The relation 
\begin{equation}   
a:\mathbb{R}^5\longrightarrow\mathbb{R}^5,\qquad a(x_1,x_2,x_3,x_4,x_5)=(x_2,x_3,x_4,x_5,x_1) 
\end{equation} 
defines an orthogonal representation of the cyclic group 
$C_5=\langle \ a\ | \ a^5=e\ \rangle =\{ e, a, a^2, a^3, a^4 \}$ 
in the usual five-dimensional Euclidean space $\mathbb{R}^5$.
The space $\mathbb{R}^5$ can be decomposed into a sum of orthogonal $C_5$-invariant subspaces 
$\mathbb{R}^5={\bf E}\oplus {\bf E}'\oplus {\bf E}''$, where
\[ 
\begin{array}{l}
{\bf E}\!=\!\left. \left\{  \alpha \left(1,{\rm cos}\frac{2\pi }{5},{\rm cos}\frac{4\pi }{5},
{\rm cos}\frac{6\pi }{5},{\rm cos}\frac{8\pi }{5}\right)\!+\! \beta 
\left(0,{\rm sin}\frac{2\pi }{5},{\rm sin}\frac{4\pi }{5}, {\rm sin}\frac{6\pi }{5},
{\rm sin}\frac{8\pi }{5}\right)\ \right|\ \alpha , \beta \!\in \!\mathbb{R} \right\} \\[2mm]
{\bf E}'\!=\!\left. \left\{ \alpha \left(1,{\rm cos}\frac{4\pi }{5},{\rm cos}\frac{8\pi }{5},
{\rm cos}\frac{2\pi }{5},{\rm cos}\frac{6\pi }{5}\right)\!+\! \beta 
\left(0,{\rm sin}\frac{4\pi }{5},{\rm sin}\frac{8\pi }{5}, {\rm sin}\frac{2\pi }{5},
{\rm sin}\frac{6\pi }{5}\right)\ \right|\ \alpha , \beta \!\in \!\mathbb{R} \right\} \\[2mm]
{\bf E}''\!=\!\{ \alpha (1,1,1,1,1)\ |\ \alpha \in \mathbb{R} \}
\end{array} 
\] 
It is known that ${\rm cos}\frac{\pi }{5}=\frac{\tau }{2}$ and
${\rm cos}\frac{2\pi }{5}=-\frac{\tau '}{2}$, 
where $\tau $, $\tau '$ are the irrational numbers $\tau =(1+\sqrt{5})/2$ and $\tau '=(1-\sqrt{5})/2$. 

The vectors $e_1=(1,0,0,0,0)$, $e_2=(0,1,0,0,0)$, ... , 
$e_5=(0,0,0,0,1)$ form the canonical basis of $\mathbb{R}^5$, and the matrices of the orthogonal 
projectors corresponding to ${\bf E}$, ${\bf E}'$ and ${\bf E}''$ in this basis are 
\begin{equation} \label{projectors} 
\begin{array}{lll}
\pi  =  \frac{1}{5}{\cal M}(2,-\tau ', -\tau ),\quad  &
\pi '  = \frac{1}{5}{\cal M}(2,-\tau , -\tau '),\quad  &     
\pi ''  = \frac{1}{5}{\cal M}(1,1,1)
\end{array}
\end{equation}
where  
\begin{equation} 
{\cal M}(\alpha ,\beta ,\gamma )=\left( \begin{array}{rrrrr} 
\alpha &\ \beta &\ \gamma &\ \gamma &\ \beta  \\ 
\beta & \ \alpha &\ \beta &\ \gamma &\ \gamma \\  
\gamma &\ \beta & \ \alpha &\ \beta &\ \gamma \\ 
\gamma &\ \gamma &\ \beta &\ \alpha &\ \beta  \\ 
\beta &\ \gamma &\ \gamma &\ \beta &\ \alpha  
	     \end{array} \right) . 
\end{equation} 
The projections on the space ${\bf E}$ of the endpoints of the vectors $e_1$, $e_2$, ... , $e_5$
are the vertices of a regular pentagon.
Any point $x\in \mathbb{Z}^5$ has ten arithmetical neighbours, namely, $x\pm e_1$, $x\pm e_2$,
$x\pm e_3$, $x\pm e_4$, $x\pm e_5$, and the projections on ${\bf E}$ of these points
are the vertices of a regular decagon. If we project on ${\bf E}$ the 2-faces of the unit hypercube $(0,1)^5$ we get,
up to certain rotations, only two rhombs.

We embed the physical two-dimensional space into $\mathbb{R}^5$ by identifying it with 
the subspace ${\bf E}$, and regard the complementary space ${\bf E}^\perp 
={\bf E}'\oplus {\bf E}''$ which corresponds to the projector $\pi ^\perp =\pi '+\pi ''$  
as an internal space. By using the strip
\begin{equation} 
{\bf S}=\{ \ x\in \mathbb{R}^5\ |\ \pi ^\perp x\in {\bf W}\ \}={\bf E}+(0,1)^5 
\end{equation} 
corresponding to the window ${\bf W}=\pi ^\perp \left((0,1)^5\right)$ we define the pattern \cite{Ka}
\begin{equation}  
\mathcal{P}=\pi \left( \mathbb{Z}^5\cap {\bf S}\right)=
\{ \ \pi x\ |\ x\in \mathbb{Z}^5,\ \pi ^\perp x\in {\bf W}\ \}. 
\end{equation} 
The window ${\bf W}$ defined as the projection on the three-dimensional space ${\bf E}^\perp $
of the open unit hypercube $(0,1)^5$ is a polyhedron and one can remark that its frontier 
contains some points belonging to $\pi ^\perp (\mathbb{Z}^5)$.
The pattern $\mathcal{P}$ is interesting, but it is a particular one. For any vector $v\in {\bf E}' $ 
we can consider the translated strip $v+{\bf S}$ which corresponds to the translated window
$v+{\bf W}$, and define the pattern
\begin{equation}  \label{pattern}
\mathcal{P}_v=\pi \left( \mathbb{Z}^5\cap (v+{\bf S})\right)=
\{ \ \pi x\ |\ x\in \mathbb{Z}^5,\ \pi ^\perp x\in v+{\bf W}\ \}. 
\end{equation} 
We say that $\mathcal{P}_v$ is a {\it non-singular
pattern} if $v$ is such that the frontier 
of $v+{\bf W}$ does not contain any element of $\pi ^\perp (\mathbb{Z}^5)$.
In the opposite case we say that $\mathcal{P}_v$ is a {\it singular pattern}.
One can prove that the points of any non-singular pattern $\mathcal{P}_v$ are the vertices of a tiling
of the plane ${\bf E}$ formed by only two tiles, each of them in a finite number of orientations \cite{Ka}.

The hyperplane
\begin{equation}  
\mathcal{E}={\bf E}\oplus {\bf E}'
=\left. \left\{ \ (x_1,x_2,x_3,x_4,x_5)\in \mathbb{R}^5\ \right| \ x_1+x_2+x_3+x_4+x_5=0 \ \right\} 
\end{equation} 
is orthogonal to the one-dimensional space 
${\bf E}''=\{ \ ( \alpha ,\alpha ,\alpha ,\alpha ,\alpha )\ |\ \alpha \in \mathbb{R} \ \}$, 
and the lattice $\mathbb{Z}^5$ is contained in the family of parallel equidistant spaces
$\{ \ \mathcal{E}_n\ |\ n\in \mathbb{Z}\ \}$, where
\begin{equation}  
\mathcal{E}_n=nw+\mathcal{E}=
\left. \left\{ \ (x_1,x_2,x_3,x_4,x_5)\in \mathbb{R}^5\ \right| \ x_1+x_2+x_3+x_4+x_5=n \ \right\} 
\end{equation} 
and $w=(0.2,0.2,0.2,0.2,0.2)$. If $x=(x_1,x_2,x_3,x_4,x_5)$ belongs to the open unit hypercube $(0,1)^5$
then $0<x_1+x_2+x_3+x_4+x_5<5$, and therefore only the hyperplanes $\mathcal{E}_1$, $\mathcal{E}_2$,
$\mathcal{E}_3$ and $\mathcal{E}_4$ meet $(0,1)^5$. This means that the set $\mathbb{Z}^5\cap (v+{\bf S})$
occurring in the definition of $\mathcal{P}_v$ is contained in 
$\mathcal{E}_1 \cup \mathcal{E}_2\cup \mathcal{E}_3\cup \mathcal{E}_4$, whence 
\begin{equation}  
\mathcal{P}_v=
\bigcup_{n=1}^4\left\{ \ \pi x\ |\ x\in \mathbb{Z}^5\cap \mathcal{E}_n , \ \pi 'x\in v+{\bf W}_n\ \right\} 
\end{equation}  
where ${\bf W}_n=\pi '\left( \mathcal{E}_n\cap (0,1)^5\right)=-nw+(\mathcal{E}_n\cap {\bf W})$.
By direct computation, one can prove that the set ${\bf W}_1$ is the interior of the regular pentagon 
with the vertices $\pi 'e_1$, $\pi 'e_3$, $\pi ' e_5$, $\pi 'e_2$, $\pi ' e_4$ (considered in this order), 
${\bf W}_2=-\tau {\bf W}_1$, ${\bf W}_3=\tau {\bf W}_1$ and ${\bf W}_4=-{\bf W}_1$.  
  
\section{Self-similarities of the singular pattern $\mathcal{P}$} 

The translation corresponding to the vector $5w=(1,1,1,1,1)\in \mathbb{Z}^5\cap {\bf E}''$
\begin{equation}  
\mathbb{Z}^5\cap \mathcal{E}_n\longrightarrow \mathbb{Z}^5\cap \mathcal{E}_{n+5}: x\mapsto x+5w 
\end{equation} 
is a one-to-one transformation, $\pi (x+5w)=\pi x$ and $\pi '(x+5w)=\pi 'x$. If we put ${\bf W}_{5k}=\emptyset $,
${\bf W}_{5k+1}={\bf W}_1$, ${\bf W}_{5k+2}=-\tau {\bf W}_1$, ${\bf W}_{5k+3}=\tau {\bf W}_1$ and 
${\bf W}_{5k+4}=-{\bf W}_1$ for any $k\in \mathbb{Z}$ then we can re-write the definition of $\mathcal{P}_v$ as
\begin{equation}  
\mathcal{P}_v=\bigcup_{n\in \mathbb{Z}}\left\{ \ \pi x\ |
\ x\in \mathbb{Z}^5\cap \mathcal{E}_n , \ \pi 'x\in v+{\bf W}_n\ \right\}. 
\end{equation} 

We say \cite{Ka,Ma} that $\lambda \in \mathbb{R}$ is an {\it inflation factor} of $\mathcal{P}_v$ if
there exists a point $y\in {\bf E}$  (called an {\it inflation center}) 
such that $\mathcal{P}_v$ is invariant under the mapping 
${\bf E}\longrightarrow {\bf E}: x\mapsto y+\lambda (x-y)$, that is,  
$x\in \mathcal{P}_v\ \Longrightarrow \ y+\lambda (x-y)\in \mathcal{P}_v $.\\[5mm]
{\bf Theorem 1. }
{\it The matrix $A=\lambda \pi +\lambda '\pi '+\lambda ''\pi ''$ has 
integer entries if and only if $(\lambda , \lambda ',\lambda '')$ belongs to the set} 
\[ \begin{array}{l} 
\mathcal{L}\!=\!\left\{  \left( \frac{ \alpha - \beta }{2}+\beta \tau ,\frac{\alpha -\beta }{2}+\beta \tau ',
\frac{\alpha +\gamma }{2}+2\gamma \right)\ ; \ \alpha ,\beta ,\gamma \in \mathbb{Z}, \ 
(-1)^\alpha \!=\!(-1)^\beta \!=\!(-1)^\gamma \ \right\}. \end{array}
\] 
{\bf Proof.}
From (\ref{projectors}) we get 
\[ \begin{array}{l}
A=\mathcal{M}\left(
\frac{2}{5}\lambda +\frac{2}{5}\lambda '+\frac{1}{5}\lambda '',
\frac{-1+\sqrt{5}}{10}\lambda +\frac{-1-\sqrt{5}}{10}\lambda '+\frac{1}{5}\lambda '',
\frac{-1-\sqrt{5}}{10}\lambda +\frac{-1+\sqrt{5}}{10}\lambda '+\frac{1}{5}\lambda ''\right).\end{array}
\]
If this matrix has integer entries then $2\lambda +2\lambda '+\lambda ''\in 5\mathbb{Z}$,
$-\lambda -\lambda ' +2\lambda ''\in 5\mathbb{Z}$ and $\lambda -\lambda '\in \mathbb{Z}\sqrt{5}$, whence
$\lambda +\lambda '\in \mathbb{Z}$. Denoting $\lambda +\lambda '=\alpha $ and $\lambda -\lambda '=\beta \sqrt{5}$
we obtain $\lambda =\frac{ \alpha - \beta }{2}+\beta \tau $, $\lambda '=\frac{\alpha -\beta }{2}+\beta \tau '$ 
and $-\alpha +2\lambda ''\in 5\mathbb{Z}$.
If we put $-\alpha +2\lambda ''=5\gamma $ then $\lambda ''=\frac{\alpha +5\gamma }{2}$ and 
$A=\mathcal{M}\left(\frac{\alpha +\gamma }{2},\frac{\gamma +\beta }{2},\frac{\gamma -\beta }{2}\right)$. 
In order to have integer entries the numbers $\alpha $, $\beta $ and $\gamma $ must have the same parity.
\qquad $\rule{2mm}{2mm} $\\[5mm]
{\bf Theorem 2.} 
{\it If $(\lambda ,\lambda ',\lambda '')$ belongs to the subset
\begin{equation} 
 \tilde{\mathcal{L}}=\{ \ (\lambda ,\lambda ',\lambda '')\in \mathcal{L}\ |\ \ 
\lambda '{\bf W}_n\subseteq {\bf W}_{\lambda ''n},\ \ for\ any\ \ n\in \{ 1,2,3,4\}\ \} 
\end{equation} 
then $\lambda $ is an inflation factor of the singular pattern $\mathcal{P}$.}\\[3mm]
{\bf Proof.}
If $(\lambda ,\lambda ',\lambda '')\in \tilde{\mathcal{L}}$ then
the corresponding mapping $A:\mathbb{R}^5\longrightarrow \mathbb{R}^5$, 
$A=\lambda \pi +\lambda '\pi '+\lambda ''\pi ''$ has the property
\begin{equation}  
\left. \begin{array}{r}
x\in \mathbb{Z}^5\cap \mathcal{E}_n \\
\pi 'x\in {\bf W}_n
\end{array} \right\} \quad \Longrightarrow \quad \left\{
\begin{array}{l}
Ax\in \mathbb{Z}^5\cap \mathcal{E}_{\lambda ''n} \\
\pi '(Ax)\in {\bf W}_{\lambda ''n}
\end{array} \right. 
\end{equation} 
In view of the definition (\ref{pattern}) considered for $v=(0,0,0,0,0)$, this means that 
$\pi x\in \mathcal{P}\Longrightarrow \pi (Ax)\in \mathcal{P}$. But 
$\pi (Ax)=\lambda \pi x$, and hence, we have $y\in \mathcal{P}\Longrightarrow \lambda y\in \mathcal{P}$.
\qquad $\rule{2mm}{2mm} $\\[5mm]
{\bf Theorem 3.}
{\it Each element of the set
\[ \begin{array}{ll}
\Lambda &=\left\{ \left. 1-\frac{\beta +5\gamma }{2} +\beta \tau \ \right| \ \beta , \gamma \in \mathbb{Z},\ \ (-1)^\beta =(-1)^\gamma ,
\ \ -\frac{\tau }{2}\leq 1-\frac{\beta +5\gamma }{2} +\beta \tau ' \leq 1 \right\} \\[2mm]
&\cup \left\{ \left. 2-\frac{\beta +5\gamma }{2} +\beta \tau \ \right| \ \beta , \gamma \in \mathbb{Z}, \ \ (-1)^\beta =(-1)^\gamma ,
\ \ -\frac{1}{2}\leq 2-\frac{\beta +5\gamma }{2} +\beta \tau ' \leq \frac{1}{\tau } \right\} \\[2mm]
&\cup \left\{ \left. 3-\frac{\beta +5\gamma }{2} +\beta \tau \ \right| \ \beta , \gamma \in \mathbb{Z}, \ \ (-1)^\beta =(-1)^\gamma ,
\ \ -\frac{1}{\tau }\leq 3-\frac{\beta +5\gamma }{2} +\beta \tau ' \leq \frac{1}{2} \right\} \\[2mm]
&\cup \left\{ \left. 4-\frac{\beta +5\gamma }{2} +\beta \tau \ \right| \ \beta , \gamma \in \mathbb{Z}, \ \ (-1)^\beta =(-1)^\gamma ,
\ \ -1\leq 4-\frac{\beta +5\gamma }{2} +\beta \tau '  \leq \frac{\tau }{2} \right\} 
\end{array}
\] 
is an inflation factor of $\mathcal{P}$.}\\[3mm]
{\bf Proof.}
It is sufficient to analyse the cases $\lambda ''\in \{ 0,1,2,3,4,5\}$. In the case
$\lambda ''=0$ we do not obtain any inflation factor. 
The ratio between the distance from the centre of a regular pentagon to one of
its sides and the distance from the centre to one vertex is cos$\frac{\pi }{5}=\frac{\tau }{2}$.
By imposing the conditions
\[ 
\begin{array}{ll}
\lambda '{\bf W}_1\subseteq {\bf W}_1, \quad \lambda '{\bf W}_2\subseteq {\bf W}_2,\quad 
\lambda '{\bf W}_3\subseteq {\bf W}_3, \quad \lambda '{\bf W}_4\subseteq {\bf W}_4&  {\rm in\ the\ case} \ \lambda ''=1,\\
\lambda '{\bf W}_1\subseteq {\bf W}_2, \quad \lambda '{\bf W}_2\subseteq {\bf W}_4,\quad 
\lambda '{\bf W}_3\subseteq {\bf W}_1, \quad \lambda '{\bf W}_4\subseteq {\bf W}_3&  {\rm in\ the\ case} \ \lambda ''=2,\\
\lambda '{\bf W}_1\subseteq {\bf W}_3, \quad \lambda '{\bf W}_2\subseteq {\bf W}_1,\quad 
\lambda '{\bf W}_3\subseteq {\bf W}_4, \quad \lambda '{\bf W}_4\subseteq {\bf W}_2&  {\rm in\ the\ case} \ \lambda ''=3,\\
\lambda '{\bf W}_1\subseteq {\bf W}_4, \quad \lambda '{\bf W}_2\subseteq {\bf W}_3,\quad 
\lambda '{\bf W}_3\subseteq {\bf W}_2, \quad \lambda '{\bf W}_4\subseteq {\bf W}_1&  {\rm in\ the\ case} \ \lambda ''=4
\end{array}
\]
we get the set $\Lambda $.\qquad $\rule{2mm}{2mm} $\\[5mm]
The well-known inflation factor $\lambda \!=\!-\tau $ belongs to $\Lambda $, and is obtained for 
$\gamma\!=\!1$, $\beta \!=\!-1$. One can remark that $\Lambda $ is a union of four cut and project sets. 
Since the elements of the multiplicative semigroup  $\mathcal{I}=\{ \, (-\tau )^k\ |\ k\in \mathbb{N}\, \}$
are not uniformly distributed in $\mathbb{R}$, we can not have $\Lambda \subset \mathcal{I}$.\\[5mm]
{\bf Theorem 4.} 
{\it If $(\lambda ,\lambda ',\lambda '')$ belongs to the subset
\begin{equation}  
\tilde{\mathcal{L}}_0=\{ \ (\lambda ,\lambda ',\lambda '')\in \mathcal{L}\ |\ \ 
\lambda '\overline{\bf W}_n\subset {\bf W}_{\lambda ''n}\ \ for\ any\ \ n\in \{ 1,2,3,4\}\ \} 
\end{equation} 
then there is an infinite 
number of inflation centers $y\in {\bf E}$ such that 
\begin{equation}  
{\bf E}\longrightarrow {\bf E}: \ x\mapsto y+\lambda (x-y)
\end{equation} 
is an affine self-similarity of $\mathcal{P}$, that is, such that  
$x\in \mathcal{P}\Longrightarrow y+\lambda (x-y)\in \mathcal{P}$.}\\[3mm]
{\bf Proof.}
If $(\lambda ,\lambda ',\lambda '')\in \tilde{\mathcal{L}}_0$ then there is a neighbourhood 
$\mathcal{V}_{(\lambda ,\lambda ',\lambda '')}$ of the origin in ${\bf E}'$ such that the relation
$(1-\lambda ')u+ \lambda '{\bf W}_n\subset {\bf W}_{\lambda ''n}$ is verified for any $n\in \{ 1,2,3,4\}$ 
and any $u\in \mathcal{V}_{(\lambda ,\lambda ',\lambda '')}$. Since the projection 
$\pi '(\mathbb{Z}^5\cap \mathcal{E})$ is dense in ${\bf E}'$, the set
$\mathcal{C}_{(\lambda ,\lambda ',\lambda '')}=\left\{ t\in \mathbb{Z}^5\cap \mathcal{E}\ \left|
\ \pi 't\in \mathcal{V}_{(\lambda ,\lambda ',\lambda '')}\right. \right\}$ 
is an infinite set. For each $t\in \mathcal{C}_{ \ (\lambda ,\lambda ',\lambda '')}$ the mapping 
$A_t:\mathbb{R}^5\longrightarrow \mathbb{R}^5$,  
$A_tx=t+(\lambda \pi +\lambda '\pi '+\lambda ''\pi '')(x-t)$  
satisfies the relations 
$\pi '(A_tx)=\pi 't+\lambda ' \pi '(x-t)=(1-\lambda')\pi 't+\lambda ' \pi 'x$, and 
\begin{equation}  
\left. \begin{array}{r}
x\in \mathbb{Z}^5\cap \mathcal{E}_n \\
\pi 'x\in {\bf W}_n
\end{array} \right\} \quad \Longrightarrow \quad \left\{
\begin{array}{l}
A_tx\in \mathbb{Z}^5\cap \mathcal{E}_{\lambda ''n} \\
\pi '(A_tx)\in {\bf W}_{\lambda ''n}
\end{array} \right. 
\end{equation} 
This means that $\pi x\in \mathcal{P}\Longrightarrow \pi (A_tx)\in \mathcal{P}$, that is,
$\pi x\in \mathcal{P}\Longrightarrow \pi t+\lambda (\pi x-\pi t)\in \mathcal{P}$.\qquad $\rule{2mm}{2mm} $\\[5mm]

\section{Self-similarities of the patterns $\mathcal{P}_v$} 

In order to avoid certain `defects' in tiling we have to translate the strip ${\bf S}$ by a vector 
$v\in {\bf E}'$ such that the frontier of $v+{\bf S}$ contains no point of $\mathbb{Z}^5$, and to
define $\mathcal{P}_v=\pi \left(\mathbb{Z}^5\cap (v+{\bf S})\right)$. The translated strip 
$v+{\bf S}$ corresponds to the translated window $v+{\bf W}=\pi ^\perp \left( v+(0,1)^5\right)$.
In this case $\pi '\left( \mathcal{E}_n\cap (v+(0,1)^5)\right)=-nw+\mathcal{E}_n\cap (v+W)=v+{\bf W}_n$,
for any $n\in \{ 1,2,3,4\}$.\\[5mm]
{\bf Theorem 5.}
{\it If $(\lambda ,\lambda ',\lambda '')\in \tilde{\mathcal{L}}_0$
then there is an infinite number of inflation centers $y\in {\bf E}$ such that 
\begin{equation} 
{\bf E}\longrightarrow {\bf E}: \ x\mapsto y+\lambda (x-y)
\end{equation}
is an affine self-similarity of $\mathcal{P}_v$, that is, such that  
$x\in \mathcal{P}_v\Longrightarrow y+\lambda (x-y)\in \mathcal{P}_v$.}\\[3mm]
{\bf Proof.}
If $(\lambda ,\lambda ',\lambda '')\in \tilde{\mathcal{L}}_0$ then there is a neighbourhood 
$\mathcal{U}_{(\lambda ,\lambda ',\lambda '')}$ of $v$ in ${\bf E}'$ such that the relation
$(1-\lambda ')(u-v)+ \lambda '{\bf W}_n\subset {\bf W}_{\lambda ''n}$ is verified for any $n\in \{ 1,2,3,4\}$ 
and any $u\in \mathcal{U}_{(\lambda ,\lambda ',\lambda '')}$. Since the projection 
$\pi '(\mathbb{Z}^5\cap \mathcal{E})$ is dense in ${\bf E}'$, the set
$\mathcal{D}_{(\lambda ,\lambda ',\lambda '')}=\left\{ t\in \mathbb{Z}^5\cap \mathcal{E}\ \left|
\ \pi 't\in \mathcal{U}_{(\lambda ,\lambda ',\lambda '')}\right. \right\}$ 
is an infinite set. For each $t\in \mathcal{D}_{ \ (\lambda ,\lambda ',\lambda '')}$ the mapping 
$A_t:\mathbb{R}^5\longrightarrow \mathbb{R}^5$,   
$A_tx=t+(\lambda \pi +\lambda '\pi '+\lambda ''\pi '')(x-t)$  
satisfies the relations 
$\pi '(A_tx)=\pi 't+\lambda ' \pi '(x-t)=v+(1-\lambda')(\pi 't-v)+\lambda ' (\pi 'x-v)$
and 
\begin{equation}  
\left. \begin{array}{r}
x\in \mathbb{Z}^5\cap \mathcal{E}_n \\
\pi 'x\in v+{\bf W}_n
\end{array} \right\} \quad \Longrightarrow \quad \left\{
\begin{array}{l}
A_tx\in \mathbb{Z}^5\cap \mathcal{E}_{\lambda ''n} \\
\pi '(A_tx)\in v+{\bf W}_{\lambda ''n}
\end{array} \right. 
\end{equation} 
This means that $\pi x\!\in \!\mathcal{P}_v\Longrightarrow \pi (A_tx)\!\in \!\mathcal{P}_v$, that is,
$\pi x\!\in \!\mathcal{P}_v\Longrightarrow \pi t\!+\!\lambda (\pi x\!-\!\pi t)\!\in \!\mathcal{P}_v$.\qquad $\rule{2mm}{2mm} $\\[5mm]

\section{Concluding remark}

In the existing models, generally, the pentagonal/decagonal quasicrystals are considered to be
two-dimensional, the third dimension being treated separately. Recently, Ben-Abraham {\it et. al.} \cite{BA}
have shown that strip projection method offers a natural way to describe these quasicrystals as three-dimensional quasicrystals. It is sufficient to choose in the above decomposition ${\bf E}^\perp $ as a physical space instead 
of ${\bf E}$. By a simple change of interpretation, our mathematical results concerning the Penrose type tilings 
become mathematical results concerning the model proposed by Ben-Abraham {\it et. al.} These new structures
are invariant under certain transformations which act as inflation but with different constants in quasiperiodic
layers and orthogonal direction.

\section*{Acknowledgment}

The author is grateful to one of the referees for some very useful suggestions.
This  research was supported by the grant CEx05-D11-03.

\end{document}